\documentclass[aps,pra,reprint,superscriptaddress,nofootinbib,longbibliography]{revtex4-1}

\usepackage{flushend}
\usepackage{booktabs,makecell}
\usepackage{graphicx}
\usepackage{xcolor}
\usepackage{dcolumn}
\usepackage{bm}
\usepackage{amsmath}
\usepackage{upgreek}
\usepackage[colorlinks,linkcolor=red,citecolor=blue,urlcolor=red]{hyperref}
\usepackage[utf8]{inputenc}
\usepackage[T1]{fontenc}
\usepackage{mathptmx}
\usepackage{ulem}
\usepackage[colorinlistoftodos]{todonotes}
\usepackage[mathlines]{lineno}
\renewcommand{\t}[1]{\mathrm{#1}}
\AtBeginDocument{%
	\newwrite\bibnotes
	\def\bibnotesext{Notes.bib}
	\immediate\openout\bibnotes=\jobname\bibnotesext
	\immediate\write\bibnotes{@CONTROL{REVTEX41Control}}
	\immediate\write\bibnotes{@CONTROL{%
			apsrev41Control,author="08",editor="1",pages="1",title="0",year="1"}}
	\if@filesw
	\immediate\write\@auxout{\string\citation{apsrev41Control}}%
	\fi
}%

\begin{document}
	
	\title{Ultralow loss torsion micropendula for chipscale gravimetry}
	
	\author{C. A. Condos}
	\affiliation{Wyant College of Optical Sciences, University of Arizona, Tucson, AZ 85721, USA}
	
	\author{J. R. Pratt}
	\affiliation{National Institute of Standards and Technology, 100 Bureau Drive, Gaithersburg, MD 20899}
	
	\author{J. Manley}
	\affiliation{Wyant College of Optical Sciences, University of Arizona, Tucson, AZ 85721, USA}
	
	\author{A. R. Agrawal}
	\affiliation{Wyant College of Optical Sciences, University of Arizona, Tucson, AZ 85721, USA}
	
	\author{S. Schlamminger}%
	\affiliation{National Institute of Standards and Technology, 100 Bureau Drive, Gaithersburg, MD 20899}
	
	\author{C. M. Pluchar}
	\affiliation{Wyant College of Optical Sciences, University of Arizona, Tucson, AZ 85721, USA}
	
	\author{D. J. Wilson}
	\affiliation{Wyant College of Optical Sciences, University of Arizona, Tucson, AZ 85721, USA}
	
	\date{\today}
	\begin{abstract}
		\textcolor{black}{We explore a new class of chipscale torsion pendula formed by Si$_3$N$_4$ nanoribbon suspensions.  Owing to their unique hierarchy of gravitational, tensile, and elastic stiffness, the devices exhibit damping rates of  $\sim 10\;\mu$Hz and parametric gravity sensitivities near that of an ideal pendulum. 
			The suspension nonlinearity can also be used to cancel the pendulum nonlinearity, paving the way towards fully isochronous, high $Q$ pendulum gravimeters. As a demonstration, we study a 0.1 mg, 32 Hz micropendulum with a damping rate of $16\;\mu$Hz, a thermal acceleration sensitivity of $2\;\t{n}g/\sqrt{\t{Hz}}$, and a parametric gravity sensitivity of $5$ Hz/$g_0$. We record Allan deviations as low as 2.5 $\mu$Hz at 100 seconds, corresponding to a bias stability of $5\times 10^{-7}g_0$. We also demonstrate a 100-fold cancellation of the pendulum nonlinearity.  In addition to inertial sensing, our devices are well suited to proposed searches for new physics exploiting low-loss micro- to milligram-scale mechanical oscillators.}
	\end{abstract}
	
	\maketitle

	Detecting gravity with small test masses is a key program for commercial technology and fundamental physics, and enjoys a rich history of innovation \cite{marson2012short}. Recent advances have enabled micro-electromechanical (MEMS) gravimeters capable of detecting the earth's tides \cite{Middlemiss2016,tang2019high,mustafazade2020vibrating}, paving the way for applications such as low-cost geophysical  surveys \cite{carbone2020newton} and inertial navigation systems \cite{el2020inertial}.  At a different extreme, micro-electro-optomechanical systems (MEOMS) have been used to detect the gravity of sub-100-mg test masses \cite{westphal2021measurement}, and to prepare mechanical oscillators \textcolor{black}{near their motional ground state \cite{delic2020cooling,xia2025motional}.  Merging these capabilities would enable proposals to search for new physics at the classical-quantum boundary \cite{bose2025massive, moore2021searching,kryhin2023distinguishable}.}
	
	In devising new strategies to miniaturize gravimeters, tradeoffs exist between accuracy and sensitivity, practicality and extensibility.  Tethered spring-mass systems are the most popular MEMS accelerometer platform because of their compatibility with wafer-scale fabrication and capacitive readout; however, gravimetry-grade devices rely on complex non-linear springs \cite{Middlemiss2016,tang2019high} \textcolor{black}{and a relatively large test mass to compensate suspension loss. Conversely, fundamental gravity experiments have driven a recent surge of interest in micro- to milligram-scale magnetically levitated test masses \cite{hofer2022high,lewandowski2021high,fuchs2024measuring,xiong2025achievement,leng2024measurement}.}  A key advantage of this approach is reduction of mechanical dissipation, enhancing sensitivity and prospects for ground state preparation \cite{streltsov2021ground}.  The practical demands of levitation however pose a severe impediment to commercialization.
	
	Here we explore an approach to chip-scale gravimetry that combines virtues of tethered and 
	levitated systems and may hold potential for both commercial and fundamental applications, based on ultralow loss, lithographically-defined torsion micropendula \cite{pratt2023nanoscale}.  Torsion balances and pendula loom large in the history of gravimetry \cite{Gillies1993}, \textcolor{black}{and simulate levitation through their partially conservative (gravitational) stiffness, leading to a natural form of dissipation dilution \cite{quinn2014bipm,cagnoli2000damping}}. 
	Moreover, pendula---arguably the first type of gravimeter \cite{richer1731observations,agnew2020time}---encode gravity into a frequency, which has distinct advantages over displacement metrology \cite{zhang2024review}. Translating these advantages to chip-scale devices requires careful consideration of the hierarchy between external and internal stresses in nanoscale suspensions \cite{shaniv2023understanding,sementilli2022nanomechanical}, and how these can been leveraged to preserve sensitivity while reducing mass.
	
	\begin{figure}[b!]
		\vspace{-2.5mm}
		\includegraphics[width=0.94\columnwidth]{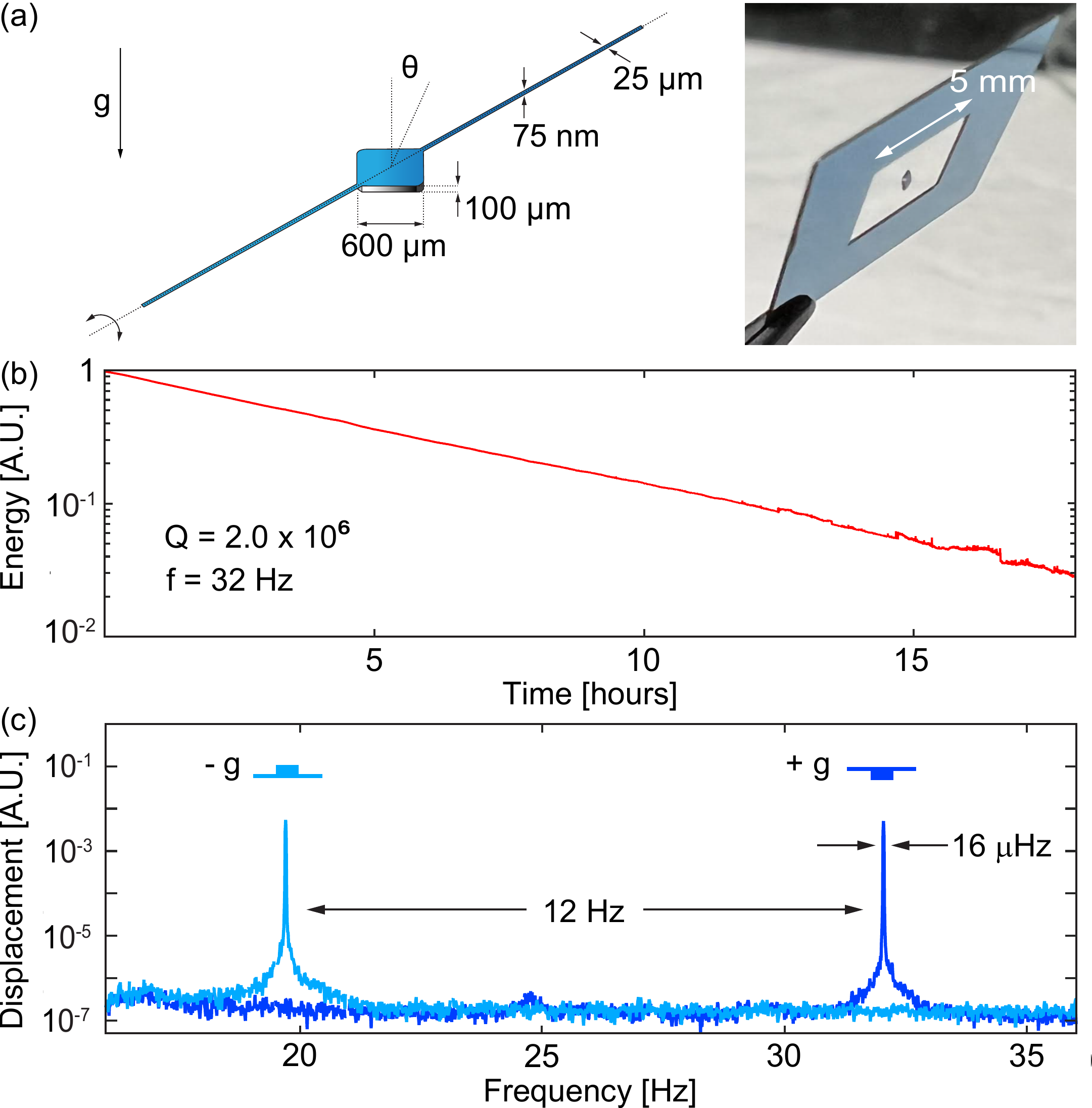}
		\caption{\textbf{Chipscale torsion micropendulum.} (a) Illustration (left) and photo (right) of a pendulum formed by suspending a 0.1 mg Si mass from a 75-nm-thick Si$_3$N$_4$ nanoribbon. (b) Energy ringdown revealing a quality factor of approximately 2 million. (c) Displacement spectrum of inverted (light blue) and normal (dark blue) device with frequency $\omega_- = 2\pi\times 20$ Hz and $\omega_+ = 2\pi\times 32$ Hz, respectively.}
		\label{fig:1}
		\vspace{-2.5mm}
	\end{figure}
	
	\begin{figure*}[ht!]
		\includegraphics[width=1.97\columnwidth]{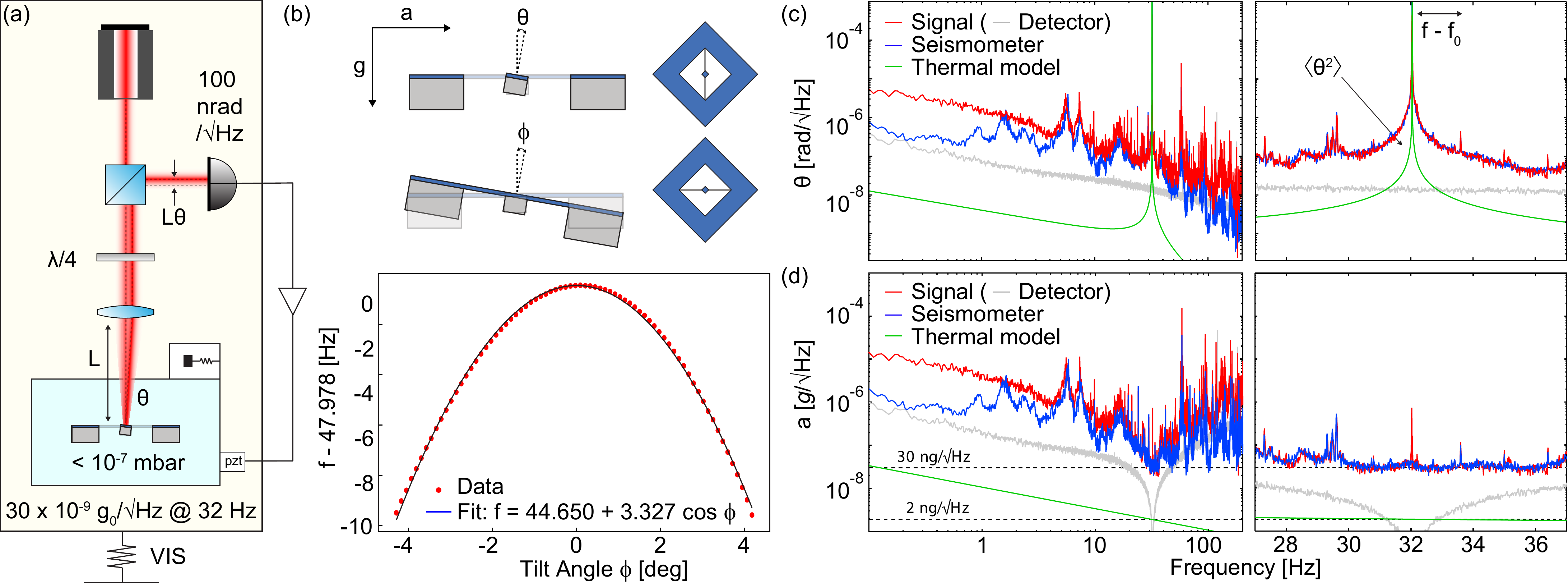}
		\caption{(a) Overview of the experiment: The micropendulum is housed in a vacuum chamber (blue) atop a vibration isolation system (VIS).  Angular displacement $\theta$ is probed with an optical lever (lever arm $L = 3\,\t{cm}$), optionally providing feedback via a piezo (``pzt").  All elements are housed in an opaque enclosure (yellow). (b) Tilt angle of pendulum ($\theta$) and device chip ($\phi$) relative to gravity $g$ and horizontal acceleration $a$ (top). Frequency versus $\phi$ for a similar device, illustrating the nonlinear stiffness of the pendulum
			(bottom). \textcolor{black}{(c) Apparent displacement and (d) acceleration of Fig. 1 device in its free-running state, compared to thermal noise models and an independent seismometer measurement.}}\label{fig:2}
		\vspace{-2mm}
	\end{figure*}
	
	A key insight of our approach, described in \cite{pratt2023nanoscale}, is that tensile stress can be used to stiffen a ribbon-like torsion suspension without adding loss. By suspending a Si microchip from a high-stress Si$_3$N$_4$ nanoribbon---as shown in Fig. \ref{fig:1}---we have realized a $0.1$ mg torsion oscillator for which the heirarchy of gravitational ($k_g$), tensile ($k_\sigma$) and elastic ($k_E$) stiffness is $k_g\approx k_\sigma \approx 10^3 k_E$.  In this regime, the oscillator's parametric sensitivity to gravity is near that of an ideal pendulum $\partial \omega/\partial g \approx \omega/2g$, while at the same time its quality factor~$(Q)$ is dramatically increased, by dissipation dilution \cite{sementilli2022nanomechanical}, to $Q \approx Q_0(k_g+k_\sigma)/k_E > 10^6$, where $Q_0^{-1}$ is the loss tangent of the suspension material. As visualized in Fig. 1c by inverting one device, the combined high sensitivity and $Q$ yields a damping-rate-equivalent (spectral) gravity resolution of $\Delta g \sim g/Q\approx 1\times10^{-6}\,g_0$ ($g_0 = 9.80665\,\t{m}/\t{s}^2$ is the standard value of acceleration due to Earth \cite{tiesinga2021codata}), an intriguing starting point for gravimetry experiments. In principle, moreover, the softening nonlinearity of gravity $k''_g < 0$ can be balanced against the hardening nonlinearity of the ribbon $k''_E>0$, yielding the prospect of an isochronous, \mbox{high-$Q$ micromechanical clock. }
	
	In this Letter, we describe a set of experiments exploring the potential of our chipscale torsion pendula as clock gravimeters.  Our near term goal is a semi-absolute gravimeter with a bias stability of $10^{-7}g_0$, sufficient to detect metrologically useful signals like the tides or an altitude change of 1 meter.  At the same time, the ability to detect gravity with a test mass near the Planck mass (22 $\upmu$g) opens up fundamental physics opportunities ranging from searches for Yukawa forces \cite{adelberger2003tests} to tests of gravitational wavefunction collapse~\cite{bassi2013models,kryhin2023distinguishable}. 
	
	The workhorse for our experiments is the device pictured in Fig. \ref{fig:1}---a replica of the device studied in \cite{pratt2023nanoscale}---consisting of a $0.6\times0.6\times 0.1\;\t{mm}^3$ 
	Si paddle suspended from a $75\;\t{nm}\times 25\;\upmu\t{m}\times 7\;\t{mm}$ Si$_3$N$_4$ nanoribbon.   The device possesses a resonance frequency, quality factor, and gravity sensitivity of $\omega_0 = 2\pi\times 32$ Hz, $Q = 2\times 10^6$, and $R = 2\pi\times 5\,\t{Hz}/g_0$, respectively, corresponding to a damping rate of $\gamma = \omega_0/Q = 2\pi\times 16\,\upmu\t{Hz}$ and a spectral resolution $\Delta g = R^{-1}\gamma = 4\times 10^{-6}g_0$.  \textcolor{black}{(A full list  of device parameters is provided in Table 1.)}  
	For all experiments, the device is housed in a vacuum chamber ($<10^{-8}\,\t{mbar}$) atop a passive vibration isolation system tuned so that it's transmissibility is minimal at $\omega_0$.  Angular displacement is recorded using a low noise optical lever~\cite{pratt2023nanoscale}.  
	
	\textcolor{black}{In the limit that its} frequency stability is thermal noise limited ($\sigma_\omega = \sqrt{\gamma/(2\epsilon  \tau)}$ \cite{sadeghi2020frequency}), the bias stability of a clock gravimeter $\sigma_g$ can be expressed as an Allan deviation (AD)
	\begin{equation}\label{eq:1}
		\sigma_g = R^{-1}\sigma_\omega = \Delta g/\sqrt{2\epsilon \gamma  \tau}
	\end{equation}
	where $\tau$ is the integration time and $\epsilon=  (\theta_0^2/2)/\langle\theta_\t{th}^2\rangle$ is the ratio of the coherent and thermal displacement power \cite{sadeghi2020frequency}.  Eq.~\ref{eq:1} suggests that for a clock gravimeter with a spectral resolution  $\Delta g = 4\times 10^{-6}g_0$,  a bias stability of $\sigma_g \approx 10^{-7}g_0$ is possible by averaging over 100 coherence times $\gamma^{-1}$; or in a single coherence time if the oscillator is driven to an amplitude 100-times in excess of thermal noise.  Leveraging these tradeoffs involves considering extraneous noise, drift, and nonlinearities, all of which can be large for nanomechanical devices.  
	
	\begin{figure*}[ht!]
		\vspace{-1mm}
		\includegraphics[width=1.97\columnwidth]{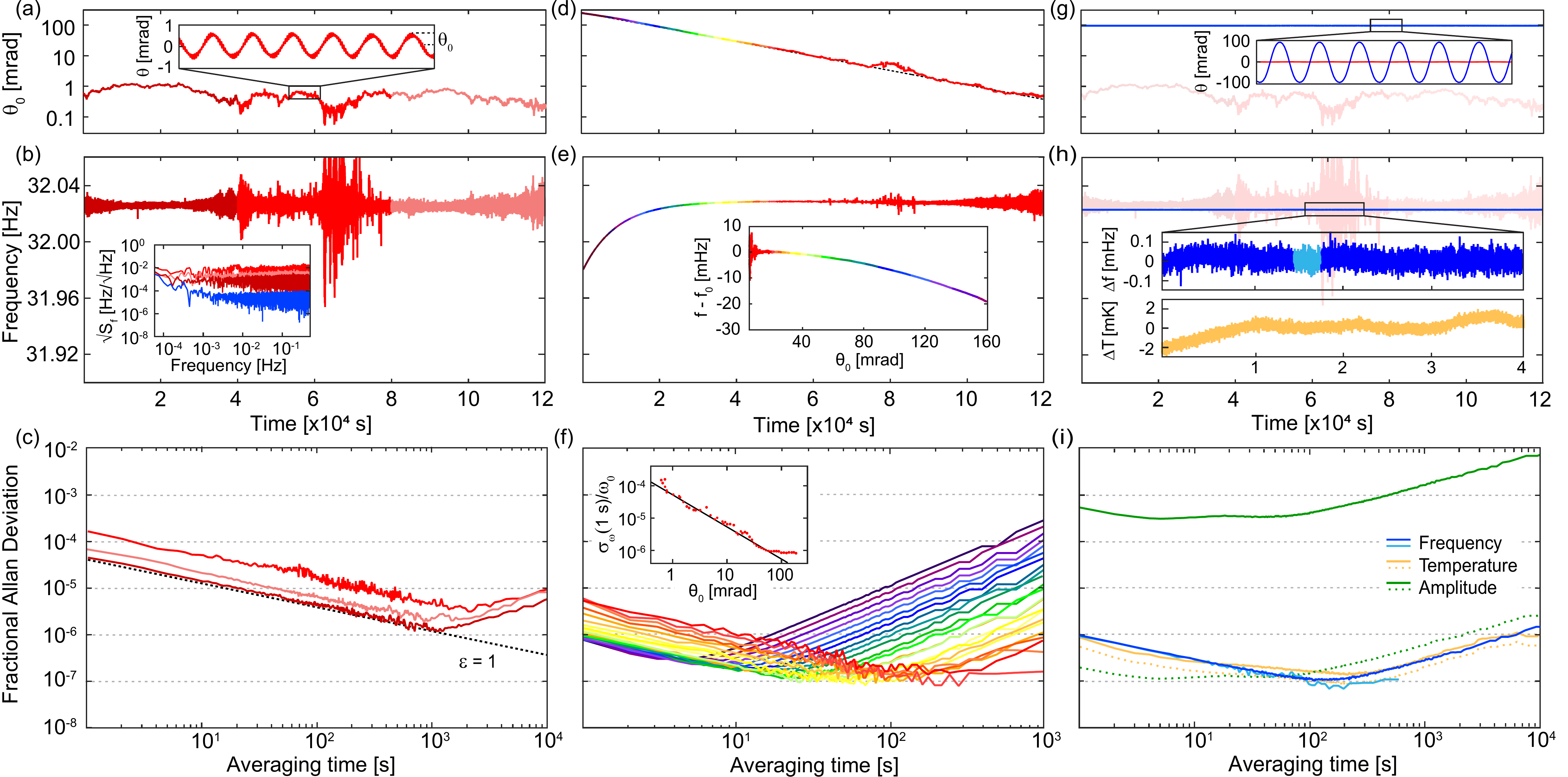}
		\caption{\textbf{Micropendulum frequency stability.} (a) Amplitude of a free-running pendulum over several hours. Inset: Displacement over several periods. (b) Corresponding frequency estimate. Inset: Frequency spectrum for highlighted intervals and data in (h) inset. (c) Fractional Allan Deviation (FAD) of data in (b). (d) Ringdown of transiently driven pendulum. (e) Frequency time series data concurrent with ringdown. Inset: Frequency shift versus amplitude. (f) FAD of successive $2000$ s segments of the data in (e). Inset: $1$-s FAD versus amplitude. (g) Amplitude of pendulum driven into self-sustained oscillation compared to free-running pendulum (red). (h) Corresponding frequency estimate. Insets: Frequency (blue) and temperature (orange) versus time over highlighted $40,000$ s interval. (i) FAD of frequency, temperature and amplitude data in (h). Dashed, colored lines are estimates of frequency-equivalent temperature and amplitude stability.}
		\label{fig:3}
		\vspace{-4mm}
	\end{figure*}
	
	As a starting point, we revisit the measurement shown in Fig. \ref{fig:1}c, in which the parametric gravity sensitivity $R$ of the pendulum is visualized by comparing its frequency in normal  $\omega_0= \omega_+$ and inverted $\omega_0= \omega_-$ chip orientations. The change in frequency reveals the fraction of the pendulum's total stiffness $k_\t{tot}\propto \omega_0^2$ which is due to gravity, which depends in turn on the direction of gravity relative to the torsion paddle, viz.
	\begin{subequations}\label{eq:2}\begin{align}
			k_\t{tot}(\phi) &= k_g \cos(\phi) + k_\sigma + k_E \\
			R(\phi) &= \frac{\partial \omega_0(\phi)}{\partial g} \textcolor{black}{\le \frac{\omega_0}{2g}\frac{k_g}{k_\t{tot}} =\frac{\omega_+^2-\omega_-^2}{4g \omega_+}
			},
	\end{align}\end{subequations}
	where $\phi$ is the chip tilt angle, defined such that $\omega_0(0) = \omega_+$ and $\omega_0(\pi) = \omega_-$ (see Fig. \ref{fig:2}b). Eq. \ref{eq:2}b allows us to directly calibrate the parametric sensitivity of the pendulum, yielding $R(0) =2\pi\times 5\,\t{Hz}/g_0$ in its normal orientation. 
	
	A key assumption behind Eq. \ref{eq:1} is that the displacement measurement is thermal noise limited---i.e., \textcolor{black}{the thermal noise spectrum
		$S_\theta^\t{th}(\Omega) = |\chi(\Omega)|^2 S_\tau^\t{th}(\Omega)$ \cite{PSD}
		exceeds readout noise at an offset frequency $\Omega-\omega_0 \approx  2\pi/\tau$, where $S_\tau^\t{th}(\Omega) = 4 k_B T I \gamma \Omega/\omega_0$, $\chi(\Omega)=I^{-1}/((\Omega^2-\omega_0^2)+i\Omega\gamma)$, and $I$ are the thermal torque, susceptibility, and moment of inertia of the pendulum, respectively \cite{komori2020attonewton,SI}.}  To investigate this requirement, as shown in Fig. \ref{fig:2}c, we probed the pendulum with a $10\;\upmu$W optical lever, and found that displacement noise dominated readout noise at $\Omega-\omega_0 \lesssim 2\pi\times 1\t{Hz}$; however, \textcolor{black}{the area beneath the noise peak $\langle \theta^2\rangle \approx (0.1\;\t{mrad})^2$ was 100-times larger than the expected thermal value $\langle \theta^2_\t{th}\rangle = k_B T/(I\omega_0^2)  \approx  (7\;\upmu\t{rad})^2$ assuming $I \approx 2\;\t{pg}\cdot\t{m}^2$ (Table 1). This excess motion 
		stems from ambient lateral acceleration at the level of $S_a(\omega_0) \approx (3\times 10^{-8} g_0/\sqrt{\t{Hz}})^2$, as evidenced in Fig. \ref{fig:2}d by comparing seismometer data to  the apparent paddle acceleration $S_a(\omega_0) = (m r_\t{CM})^2 |\chi(\Omega)|^{-2}S_\theta(\Omega)$, where  
		$m \approx 67\;\upmu\t{g}$ and $r_\t{CM}\approx 48\,\upmu$m are the paddle mass and lever arm, respectively. This discrepancy  underscores the low thermal acceleration of the micropendulum, $S_a^\t{th}(\omega_0) = 4 k_B T \gamma I /(m r_\t{CM})^2 \approx (2\times 10^{-9}g_0/\sqrt{\t{Hz}})^2$,} and implies that thermal displacement noise in Eq.~\ref{eq:1} must be replaced by $\langle \theta^2\rangle \sim 10^2\langle\theta_\t{th}^2\rangle$, while $\tau^{-1/2}$ scaling should occur for averaging times $\tau\gtrsim 1\;\t{s}$.{}
	
	Armed with these expectations, we conducted a series of experiments to determine the frequency stability of the micropendulum in Fig. \ref{fig:1}.  For these experiments, as illustrated in Fig. \ref{fig:2}a, the apparatus was housed in an opaque enclosure and the device platform was temperature-stabilized to within approximately $1\;\t{mK}$ using a Peltier cooler.  The amplitude of the pendulum was controlled using active feedback of the optical lever photocurrent to a piezo affixed to the vacuum chamber.  The frequency of the pendulum was estimated using a software fitting routine applied to the digitized photocurrent \cite{pratt2023nanoscale}. 
	
	As shown in Fig. \ref{fig:3}(a-c), we first recorded the frequency of the pendulum in its free running state, sampling at 100 kHz for 1 second intervals. At short times, the frequency Allan deviation $\sigma_\omega$ was found to scale in qualitative agreement with the thermal noise model using $\epsilon = 1$, consistent with the findings of \cite{wang2021fundamental,wang2020frequency}.  
	Beyond $10^3$ seconds, thermal noise $\sigma_\omega\propto 1/\sqrt{\tau}$ is overcome by linear drift $\sigma_\omega\propto \tau$, 
	yielding a minimum frequency stability of $\sigma_\omega\approx 1.1\times 10^{-6} \omega_0$ at $\tau \approx 1500$ s, corresponding to a gravity resolution of $\sigma_g \approx 7.0\times 10^{-6} g_0$.
	
	According to Eq. \ref{eq:1}, frequency stablity can be increased by driving the pendulum into coherent oscillation, $\epsilon > 1$, in principle realizing $\sigma_g =  10^{-7}g_0$ within the coherence time $\gamma^{-1}\approx 10^4\,\t{s}$, if the pendulum nonlinearity  is sufficiently low.  To explore this possibility, we conducted an experiment in which the pendulum was transiently excited---using feedback as shown in Fig. \ref{fig:2}---and then allowed to undergo free decay, $\theta_0(t) = \theta_1 e^{-\gamma t/2}$, resulting in a frequency ringdown~\cite{SI}
	\begin{equation}\label{eq:3}
		\omega(t) \approx \omega_0\left(1 + \frac{k''_g + k''_E}{16 k_\t{tot}}\theta_1^2 e^{-2\gamma t}\right) \equiv  \omega_0(1+\alpha_\t{NL}\theta_0^2(t))
	\end{equation}
	where $k_g'' = -k_g$ and $k_E''$ are the pendulum nonlinearity and suspension 
	nonlinearity, respectively, and $k_\t{tot}\approx k_\sigma+k_g$ is the total pendulum stiffness.  For the measurement shown in Fig. \ref{fig:3}e, the oscillator was driven to $\theta_1 = 160\,\t{mrad}$ and $\sigma_\omega$ was computed on $\tau=10^3$ s intervals.  Evidently, driving produces the anticipated reduction of $\sigma_\omega$ at short times; however, drift due to the pendulum nonlinearity $\alpha_\t{NL} \approx  -0.02/\t{rad}^2$ fundamentally limits the frequency stability to $\sigma_\omega\approx 2\times 10^{-7}\omega_0$ at $\tau\approx 10^2$ s, corresponding to $\sigma_g = 1.3\times 10^{-6}g_0$.  
	
	To mitigate nonlinear frequency drift, we turned our attention to stabilizing the amplitude of the micropendulum.  To this end, following a standard approach, we implemented a self-sustained oscillator by combining piezoelectric feedback with a gain-clamped amplifier.  Shown in Fig. \ref{fig:3}(g-i) is an experiment for which, after a settling time of several hours, the amplitude was stabilized to within $1\%$ of $\theta_0 \approx 100$ mrad.  The frequency, amplitude, and temperature of the pendulum were then tracked for a day. Over the highlighted 11-hour period, we observed that the frequency stability of the pendulum was limited to $\sigma_{\omega}\approx1.1\times 10^{-7}\omega_0$ at 100 s, corresponding to $\sigma_g \approx 7\times 10^{-7}g_0$.  Overlays of the temperature and amplitude AD scaled by the measured temperature sensivitity $f^{-1}\partial f/\partial T\approx 4\times 10^{-6}/\t{mK}$ \cite{SI} and amplitude sensitivity $f^{-1}\partial f/\partial\theta_0 = 2\alpha_\t{NL}\theta_0 \approx 2\times 10^{-6}/\t{mrad}$  suggest that temperature ($\sigma_T\gtrsim 30\;\upmu\t{K}$) and amplitude noise ($\sigma_{\theta_0}\gtrsim 40\;\upmu\t{rad}$) contribute equally to the observed instability.  Over the highlighted, relatively stable, 20 min interval, we observe minima $\sigma_{\omega}\approx7\times 10^{-8}\omega_0$ and $\sigma_g \approx 5\times 10^{-7}g_0$ at 200 s.

	 Our experiments suggest that the frequency stability of our micropendulum is limited by environmental noise and drift---in particular, acceleration noise, \textcolor{black}{temperature, and amplitude drift---limiting drive powers, frequency stability, and gravity resolution to $\epsilon\sim 10^4$, $\sigma_\omega\sim 2\pi\cdot 1\,\upmu\t{Hz}$, and $\sigma_g\sim 10^{-6}g_0$}, respectively. They also highlight a fundamental tradeoff, in that leveraging high drive powers entails increased sensitivity to amplitude fluctuations due to the pendulum nonlinearity:
	\begin{equation}\label{eq:4}
		\frac{\sigma_\omega^2(\tau)}{\omega_0^2} \gtrsim \frac{\langle \theta_\t{th}^2 \rangle \gamma }{\theta_0^2\tau \omega_0^2}+4\alpha_\t{NL}^2\theta_0^2\sigma^2_{\theta_0}(\tau)\ge 4\alpha_\t{NL}\sqrt{\tfrac{\sigma^2_{\theta_0} (\tau)\langle\theta_\t{th}^2\rangle}{Q\omega_0\tau }}.
	\end{equation}
	
	The nonlinear instability given by Eq. \ref{eq:4} is a well-known limitation of pendulum clocks dating back to Huygens \cite{marson2012short}, and poses a fundamental obstacle to pendulum-based sensors. However, the hierarchy of stiffnesses of our micropendula yields a surprising workaround, in that the gravitational nonlinearity $k_g'' = - k_g$ is comparable but opposite in sign to the suspension nonlinearity $k_E''$, enabling cancellation of the net nonlinearity by tailoring the ribbon dimensions \cite{pratt2023intersection}. Viz.,
	\begin{equation}
		\alpha_\t{NL} = \frac{- k_g}{16 k_\t{tot}}\left(1-\frac{k_E''}{k_g}\right)\approx \alpha_{\t{NL}}^{(g)}\left(1-\frac{3 h w^5 E}{5 L^2 mgr_\t{CM} }\right)\label{eq:5}
	\end{equation}
	where $\alpha_\t{NL}^{(g)} = -k_g/(16 k_\t{tot}$) is the gravitational nonlinearity ($-1/16$ for an ideal pendulum, $k_\t{tot} = k_g \textcolor{black}{= mg r_\t{CM}}$) and $k_E'' \approx 3E h w^5/(5 L^3)$ is the nonlinear torsional stiffness of a ribbon of thickness $h$, width $w$, and length $L$ \cite{pratt2023intersection,SI}. Indeed, noting that the ribbon tensile stiffness \textcolor{black}{$k_\sigma = \sigma h w^3/3L$} and gravitational sensitivity $R\propto (1+k_\sigma/k_g)^{-1/2}$, Eq. \ref{eq:5} implies that $\alpha_\t{NL}$ can be canceled without altering $R$, by tailoring $w$ and $r_\t{CM}$.

	\begin{figure}[t!]
		\includegraphics[width=0.95\columnwidth]{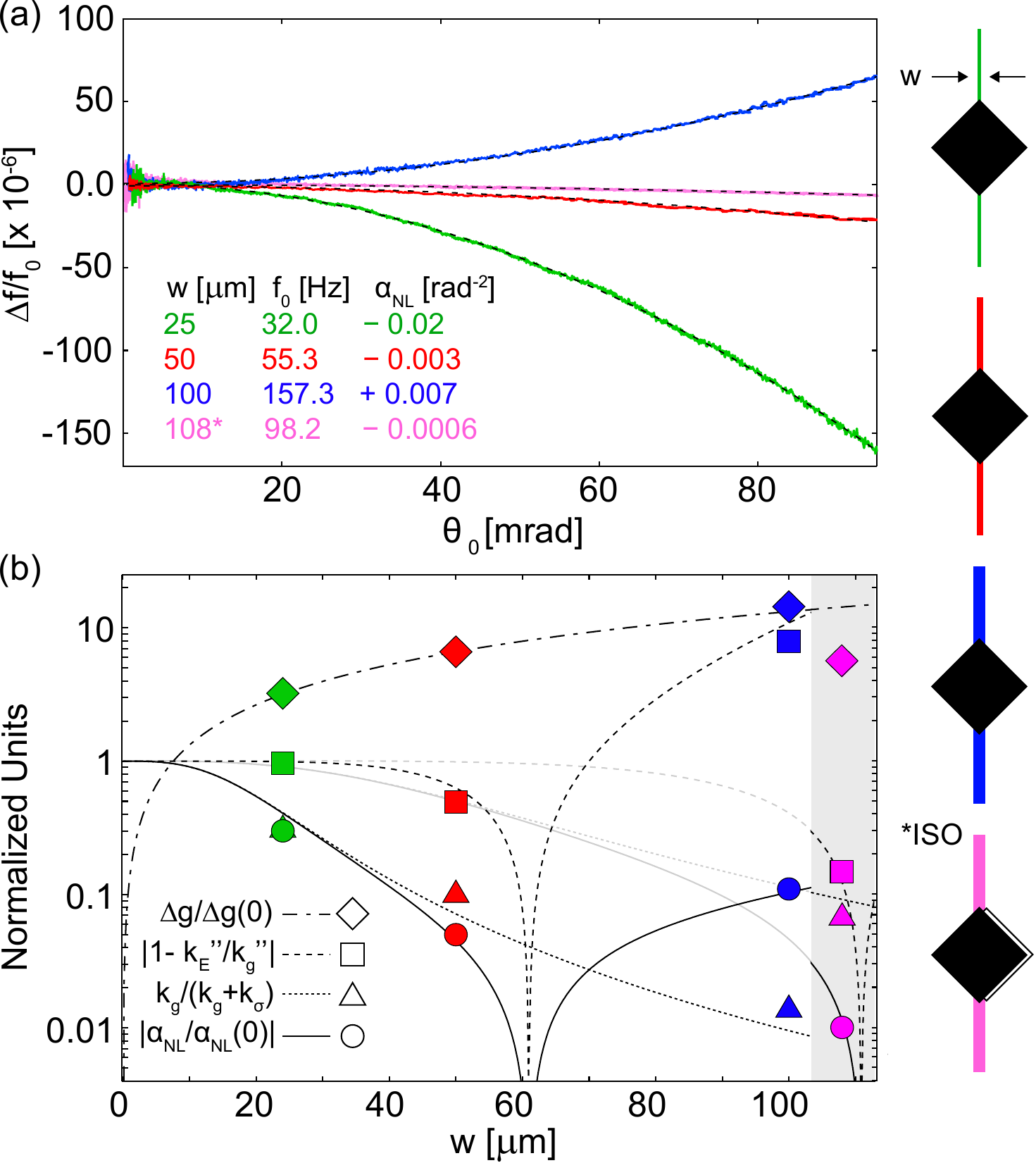}
		\caption{\textbf{Towards an isochronous micropendulum.} \textcolor{black}{(a) Frequency versus amplitude for micropendula with different suspension widths. Dashed lines are fits to Eq. \ref{eq:3}. (b) Measurement (markers) and models (lines) of frequency nonlinearity $\alpha_\t{NL}$, damping-equivalent gravity resolution $\Delta g = R^{-1} f/Q$, participation factor of gravitational stiffness $k_g/k_\t{tot}$, and relative nonlinearity $|(k_g - k_E'')/k_g|$, assuming the paddle geometry in Fig. \ref{fig:1}a. $\alpha_\t{NL}(w=0)$ and $\Delta g (w=0)$ are for an ideal pendulum. The magenta ``isochronous'' (ISO) device uses a thicker paddle and suspension width fine tuned to minimize $\alpha_\t{NL}$ while preserving $R$ and $\Delta g$. 
				See Table 1 and the SI for details.}}
		\label{fig:4}
		\vspace{-2mm}
	\end{figure}

	To explore the possibility of canceling the pendulum nonlinearity, we fabricated two additional pendula with suspension widths of $w=50\,\upmu \t{m}$ and $100\,\upmu \t{m}$ (otherwise the pad dimensions were fixed), and compared their nonlinear ringdowns and gravity sensitivity to the $w = 25\,\upmu\t{m}$ device.  \textcolor{black}{We also explored a new deep Si-etching technique (see SI) that allowed us to engineer a four times thicker (400 $\mu$m) pad with precise sidewalls.  As shown in Fig. \ref{fig:4}a, $\alpha_\t{NL}$ was observed to decrease 10-fold for the $w=50\;\upmu$m device and change sign for the $w=100\;\upmu$m device---consistent with Eq. \ref{eq:5} using $r_\t{CM} = 48\;\upmu\t{m}$ and $m = 67\,\upmu\t{g}$---while the gravity sensitivity $\Delta g$ exhibited the expected $w^{-3/2}$ scaling.  For the 400-$\mu$m-thick pad, choosing $w \approx 100\;\mu\t{m}$ yielded a near-isochronous micropendulum with $R = 3.5\;\t{Hz}/g_0$, $Q = 5.0\times 10^6$, $\Delta g = 1.0\times 10^{-6}g_0$ and $\alpha_\t{NL} \approx -0.0006/\t{rad}^2$, a 30-fold reduction over the device studied in Figs. \ref{fig:1}-\ref{fig:3}. \mbox{(See appendix for further details.)}}

	Looking forward, we envision a combination of improved environmental control and isochronicity to achieve a long-term gravity resolution less than $10^{-7}\,g_0$, sufficient for clock-based altimetry and tidal monitoring. A key challenge is thermal drift. 
	To this end, we  note that the observed frequency-temperature coefficient of our device \textcolor{black}{(4 ppm/K)} is 10-times larger than expected due to stress relaxation ($\alpha_\t{rel}E/(2\sigma)$, where $\alpha_\t{rel}$ is the relative thermal expansion coefficient of Si and Si$_3$N$_4$ \cite{wang2020frequency}), implying a \textcolor{black}{$10^2$-$10^3$ fold margin for improvement using strained Si ($\alpha_\t{rel} =0$) \cite{beccari2022strained} or thermally-invariant strain-engineering \cite{wang2020frequency}. We also note that the reduced gravity sensitivity of higher order pendulum and suspension modes allows them to be used as in-situ thermometers, a standard technique for parametric MEMS sensors \cite{zhang2024review}. }
	
	\textcolor{black}{Finally, we wish to emphasize that the low thermal acceleration and torque of our micropendula---$2\times 10^{-9}g_0/\sqrt{\t{Hz}}$ and $6\times 10^{-20}\,\t{Nm}/\sqrt{\t{Hz}}$, for the studied device---
		in conjunction with precise dimensional control, scalability, and functionalizability afforded by lithographic patterning; makes them appealing as a platform for next-generation fundamental physics experiments, including fifth-force \cite{adelberger2003tests} and dark matter \cite{carney2021mechanical} searches, spontaneous wavefunction collapse tests \cite{bassi2013models}, and quantum gravity tests \cite{rademacher2020quantum,bose2025massive}. 
		Recently we proposed a search for short-range Yukawa forces in the little-explored 10-100 $\mu$m range, for example, by suspending a torsion micropendulum above a materially heterogeneous source mass \cite{manley2024microscale}.  Conversely, heterogeneous torsion micropendula \cite{sun2024differential}, in conjunction with cryogenic operation, offer a route to accelerometer-based ultralight dark matter searches \cite{carney2021ultralight,manley2021searching} at acoustic frequencies that meets stringent sensitivity requirements \cite{manley2021searching} and is  compatible with large arrays. For continuous spontaneous wave-function localization tests \cite{diosi2015testing}, a rigid rotor with large dimensions and low torque noise is attractive \cite{schrinski2017collapse,carlesso2018non,komori2020attonewton}, and preliminary estimates suggest that a 4 K version of our device can access unexplored parameter space \cite{pandurangi2022optomechanics}. Finally, a recent proposal to distinguish classical and quantum gravity \cite{kryhin2023distinguishable} looks for correlations between a pair of high-Q mechanical oscillators produced by their mutual gravitation.  This protocol imposes similar design constraints as our short-range gravity proposal \cite{manley2024microscale}, but with an added imperative of high coherence $Q\omega_0/(k_B T/\hbar)$, which might benefit from wider suspensions through the scaling law $Q\omega_0\propto  w^{3.5}$ \cite{pratt2023nanoscale,shin2024laser}. 
		In the appendix we provide a table that summarizes key features of the reported devices and their strong geometry dependence, which can be tailored towards desired applications.}\\ 

\noindent\textit{Acknowledgments} - The authors thank Will Terrano for useful discussions and Andrew Land for experimental support. This work was supported by National Science Foundation award nos. 2239735 and 2330310. CAC acknowledges support from a UArizona National Labs Partnership Grant. ARA was supported by a CNRS-UArizona iGlobes Fellowship. 

\bibliography{ref}

\begin{thebibliography}{50}
\bibitem [{PSD()}]{PSD}%
  \BibitemOpen
  \href@noop {} {\bibinfo  {journal} {Here $S_\theta(\Omega)$ is the single-sided power spectral density of $\theta$ at Fourier frequency $\Omega$}\ }\BibitemShut {NoStop}%
\end{thebibliography}

\color{black}

\vspace{-3mm}\section*{End Matter}

\renewcommand{\arraystretch}{1.1}
\begin{table}[b!]
	\centering\color{black}
	\begin{tabular}[t]{llllllll}
		\Xhline{2\arrayrulewidth}
		& & & & \hspace{-16mm}Model (measurement)\\
		\cline{3-6}
		&Formula & Dev. 1&  Dev. 2& Dev. 3& ISO& Units&\\
		\hline
		$m$ & $\rho w_\t{p}^2 h_\t{p}$ & 67 & 67 & 67 & 336 &$\upmu\t{g}$\\
		$I$ & $\frac{m(w_p^2+4 h_p^2)}{12}$ & 1.9 & 1.9 & 1.9 & 28 &$\t{pg}\cdot \t{m}^2$\\
		$k_g$ & $m g_0 h_\t{p}/2$ & 31 & 31 & 31 & 658 &$\frac{\t{pN}\cdot\t{m}}{\t{rad}}$\\
		$k_\sigma$ & $\sigma h w^3/3L$ & 45 & 357 & 2900 & 6480 &\\
		$k_E$ & $2 E h^3 w/3L$ & 0.25 & 0.5 & 1.0 & 1.9 &\\
		$k''_E$ & $3 E h w^5/5L^3$ & 0.32 &10  & 328 & 578 &$\frac{\t{pN}\cdot\t{m}}{\t{rad}^3}$\\
		$\omega_0$ & $\sqrt{\frac{k_\sigma+k_g+k_E}{I}}$ & 32 (32) & 72 (55) & 197 (157)& 80 (98)& $2\pi\cdot\t{Hz}$\\
		$\alpha_\t{NL}$ & $\frac{1}{16}\frac{k''_E - k_g}{k_\sigma+k_g+k_E}$ & -25 (20) & -3.3 (3) & 6.5 (6) & -0.7 (0.6)&$\frac{\t{ppb}}{\t{mrad}^{2}}$\\
		$Q$ & $Q_0 \frac{k_\sigma+k_g+k_E}{k_E}$ & 1.7 (2) & 4.0 (3) & 15 (10)& 24 (5) &$10^6$\\
		$R$ & $\frac{\omega_0}{2g} \frac{k_g}{k_\sigma+k_g+k_E}$& 6.3 (5.0) & 2.9 (2.8) & 1.1 (1.1) & 3.7 (3.5) &$2\pi\cdot\frac{\t{Hz}}{g_0}$\\
		$\Delta g$ & $\frac{2g_0}{Q_0}\frac{k_E}{k_g}$& 3.1 (3.2) & 6 (6.5) & 12 (14) & 1.0 (5.6) &$10^{-6}g_0$\\
		$S_\tau^\t{th}$ & $\sqrt{4 k_B T I\tfrac{\omega_0}{Q}}$& 63 & 59 & 50 & 99 &$\frac{\t{zN}\cdot{\t{m}}}{\sqrt{\t{Hz}}}$\\
		$S_a^\t{th}\;\;\;$ & $\tfrac{2}{m h_\t{p}}\sqrt{S_\tau^\t{th}}\;\;$ & 2 (2) & 1.9 & 1.6 & 0.15 &$\frac{\t{n}g_0}{\sqrt{\t{Hz}}}$\vspace{0.5mm}\\
		\Xhline{2\arrayrulewidth}
	\end{tabular}
	\caption{\textbf{Model and measured device parameters.} Device 1 is the torsion micropendulum studied in Figs. \ref{fig:1}-\ref{fig:3}, with suspension width $w = 25\;\upmu\t{m}$. Device 2 (3) is the $w = 50\; (100)\;\upmu\t{m}$ variant in Fig. 4.  Models of devices 1-3 in Fig. 2(c,d) and 4(b) assume $\{h,L,w_\t{p},h_\t{p}\}=\{0.075,\,7000,\,550,\,95\}\;\upmu\t{m}$, where $w_\t{p}\,(h_\t{p})$ is the paddle width (thickness). ISO is the magenta device in Fig. 4, with model parameters $\{w,h,L,w_p,h_p\}=\{108,0.09,7000,600,400\}\;\upmu\t{m}$. For device 1-3 (ISO) models, we assume a suspension stress of $\sigma = 0.8\;(1.2)$ GPa. For all models, we assume Si density $\rho = 2330\;\t{kg}/\t{m}^3$, ribbon elastic modulus $E = 250\;\t{GPa}$, and intrinsic $Q_0(h) = 7000\cdot (h/\t{nm})$~\cite{villanueva2014evidence}.}
	\vspace{0mm}\end{table}

\subsection*{Geometry-dependent device properties} 

Linear and nonlinear properties of our torsion micropendula are highly dependent on the geometry of the ribbon suspension and the torsion paddle, through the functional forms of $k_{E}$, $k_\sigma$, $k_g$, $k_{E}''$, $k_{\sigma}''$, $I$, $m$, and~$r_\t{CM}$. Table I provides an overview, focusing on a diamond-shaped paddle as shown in Fig. 1(a), with width $w_\t{p}$ and thickness $h_\t{p} = 2 r_\t{CM}$.  

Note that we focus on devices for which $k_\sigma\sim k_g \gg k_E$, in which case the parametric gravity sensitivity $R = \partial \omega_0/\partial g =  (\omega_0/2g)/(1+k_\sigma/k_g)$ approximates that of an ideal pendulum $(k_\sigma =0)$.  Increasing the paddle size reinforces this behavior; however, its effect on $R$ (and $S_\tau$ and $S_a$) in general depends on paddle geometry. An alternative sensitivity metric with takes on a simplified form is the gravity-equivalent damping rate
\begin{equation}\label{eq:Delta g}
	\Delta g\equiv \frac{\omega_0}{Q}R^{-1} = \frac{2g}{Q_0}\frac{k_E}{k_g} \ge \frac{2}{Q_0}\frac{k_E}{m r_\t{CM}}
\end{equation}
Equation \ref{eq:Delta g} suggests---and Table 1 bears out---that increasing paddle mass always decreases $\Delta g$, as long as $r_\t{CM}$ is not decreased.  Note, however, that for a sufficiently large paddle that $k_g > k_\sigma$, the device can no longer be inverted.  

An extended version of Table 1 with further examples is provided in the Supporting Information (SI).

\vspace{-2mm}
\subsection*{Isochronous Micropendulum}
\vspace{-2mm}

\begin{figure}[h!]
	\includegraphics[width=0.85\columnwidth]{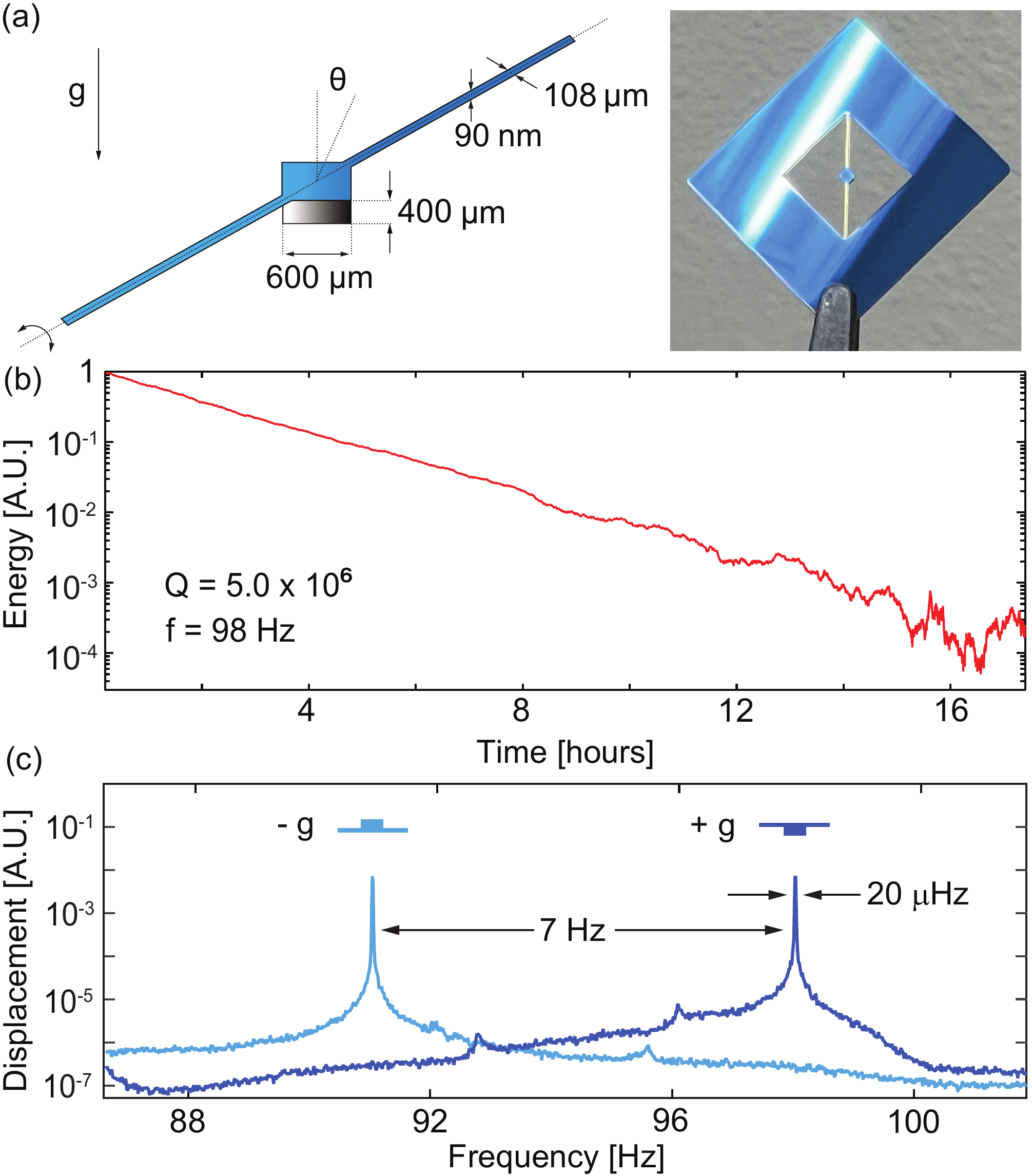}
	\caption{\color{black}\textbf{Isochronous micropendulum} (a) Illustration (left) and photo (right) of micropendulum formed by suspending a 0.3 mg Si mass from a 100-nm-thick Si$_3$N$_4$ nanoribbon. (b) Energy ringdown revealing a quality factor of approximately 5 million. (c) Displacement spectrum of device in inverted (light blue) and normal (dark blue) configuration, revealing resonance frequencies of $\omega_- = 2\pi\times 91$ Hz and $\omega_+ = 2\pi\times 98$ Hz, respectively. } 
	\label{fig:S4}
	\vspace{0mm}
\end{figure}

Geometry and characterization of the isochronous (``ISO'') micropendulum in Fig. \ref{fig:4} is presented in Fig. \ref{fig:S4}. In order to decrease the frequency nonlinearity $\alpha_\t{NL}$ while maintaining gravity sensitivity $R$, the suspension of the ISO device is widened to $w \approx 100\;\mu$m and the paddle thickness is increased to $h_\t{p} = 400\,\upmu$m. Precise control of the paddle width is necessary to achieve the desired $k_E''$.  This is achieved using a deep reactive-ion etching procedure described in the SI.  The paddle thickness is meanwhile constrained by patterning both sides of the wafer.  The less precise, single-sided wet etch used to fabricate devices 1-3 in Fig. \ref{fig:4} is described in the SI of \cite{pratt2023intersection}.

\newpage

\end{document}



\title{Supplementary information for ``Ultralow loss torsion micropendula for chipscale gravimetry''}

	\author{C. A. Condos}
	\affiliation{Wyant College of Optical Sciences, University of Arizona, Tucson, AZ 85721, USA}
	
	\author{J. R. Pratt}
	\affiliation{National Institute of Standards and Technology, 100 Bureau Drive, Gaithersburg, MD 20899}
	
	\author{J. Manley}
	\affiliation{Wyant College of Optical Sciences, University of Arizona, Tucson, AZ 85721, USA}
	
	\author{A. R. Agrawal}
	\affiliation{Wyant College of Optical Sciences, University of Arizona, Tucson, AZ 85721, USA}
	
	\author{S. Schlamminger}%
	\affiliation{National Institute of Standards and Technology, 100 Bureau Drive, Gaithersburg, MD 20899}
	
	\author{C. M. Pluchar}
	\affiliation{Wyant College of Optical Sciences, University of Arizona, Tucson, AZ 85721, USA}
	
	\author{D. J. Wilson}
	\affiliation{Wyant College of Optical Sciences, University of Arizona, Tucson, AZ 85721, USA}

\date{\today}

\begin{abstract}
Below we derive key equations in the main text, provide additional device images and details about the measurements shown in Figs 1-4, and comprehensive overview of device properties and their geometry dependence.
\end{abstract}

\maketitle

\tableofcontents

\section{Nonlinear Dynamics of a Torsion pendulum}
Consider a simple pendulum of mass $m$ and length $r$ which is constrained to oscillate in a plane at tilted $\phi$ from the gravitation field (such that $\phi=0$ for a standard pendulum and $\phi=\pi$ for an inverted pendulum).  When displaced by $\theta$ from rest, the pendulum experiences a nonlinear restoring torque
\begin{equation}
    \tau_g(\theta) = mgr\cos \phi\sin \theta
\end{equation}
or equivalently, a nonlinear stiffness
\begin{equation}
        k_g(\theta) \equiv \tau'_g(\theta) = mgr\cos \phi\cos \theta
\end{equation}
where $g$ is the local acceleration due to gravity.

Expanding about $\theta=0$ to third order gives
\begin{equation}
    \tau(\theta)\approx -k_g\theta -k_g'' \theta^3/6
\end{equation}
where for notational simplicity we define
\begin{equation}
    \begin{aligned}
        k_g&=k_g(0) = mg r\cos \phi \\
        k''_g&=k''_g(0) = -k_g
    \end{aligned}
\end{equation}
as the linear spring constant ($k_g$) and second order (Duffing) nonlinearity ($k''_g$) of the pendulum, respectively. 

Now consider a torsion pendulum which is subject to an additional restoring torque from its suspension. For a ribbon suspension with tensile stress $\sigma$ and elastic modulus $E$, the additional linear and nonlinear torsional stiffinesses are~\cite{pratt2023intersection}
\begin{equation}
    \begin{aligned}
        k_\sigma &= \frac{\sigma {h} {w}^3}{3 L}  \\
        k''_\sigma&= -\frac{3 \sigma h {w}^5}{5 {L}^3}\\
        k_E &= \frac{2 E {h}^3 w}{3 L} \\
        k''_E&= \frac{3 E h {w}^5}{5 {L}^3}
    \end{aligned}
\end{equation}
where $w$, $L$, $h$ are the ribbon width, length, and thickness, respectively. For the torsion micropendula presented in the main text, $\left|k''_\sigma \right| \ll \left|k''_E\right|$ and $k_E\ll k_\sigma$.  The net (total) linear and nonlinear torsional stiffnesses can therefore be expressed as 
\begin{equation}\label{eq:ktot}
    \begin{aligned}
        k_\t{tot} &= k_\sigma+k_E+ k_g\approx k_\sigma +k_g \\
        k''_\t{tot} &= k''_\sigma+k''_E+ k''_g\approx k''_E +k''_g 
    \end{aligned}
\end{equation}

\subsection{Frequency-amplitude coupling}
The torsion micropendula presented in the main text can be modeled with the nonlinear (Duffing) equation of motion 
\begin{equation}\label{eq:Duffing}
	I\ddot{\theta}+k_\text{tot} \theta+\frac{k''_\text{tot}}{6}\theta^3= 0
\end{equation}
where $I$ is the moment of inertia of the torsion paddle and $k_\t{tot}$ and $k''_\t{tot}$ are as given by Eq. \ref{eq:ktot}.

The solution to Eq. \ref{eq:Duffing} is well known \cite{rand2012lecture}. For a sufficiently small initial amplitude $\theta_0$, a stable orbit with a period
\begin{equation}
	T(\theta_0) \approx  \frac{2\pi}{k_\t{tot}/I}\left(1-\frac{3}{8}\frac{\beta}{\omega_1^2} \theta_0^2\right)
\end{equation}
is described, corresponding to an effective frequency
\begin{equation} \label{eq:freqShift}
        \omega(\theta_0)\equiv \frac{2\pi}{T(\theta_0)}\approx \omega_0\left(1+\alpha_\text{NL} \theta_0^2\right) \\
\end{equation}
where $\omega_0 =\sqrt{k_\text{tot}/I}$ is the ``cold'' frequency and
\begin{equation}\label{eq:alphaNL}
            \alpha_\text{NL}\equiv \frac{1}{16} \frac{k''_\text{tot}}{k_\text{tot}}.
\end{equation}
Equation \ref{eq:alphaNL} describes the frequency-amplitude coupling of the torsion pendulum.  It vanishes when $k''_\text{tot}=0$ (i.e., when $k_E + k_\sigma = -k_g$), a condition known as ``isochronous.'' 

\subsection{Parametric sensitivity to gravity}
Because the magnitude of $\tau_g$ depends on local gravitational acceleration $g$, the pendulum oscillation frequency is sensitive to changes in $g$. The sensitivity is given by
\begin{equation}\label{eq:R}
    R\equiv \frac{\partial \omega_0}{\partial g}= 
    \frac{\omega_0}{2g}\frac{k_g}{k_\t{tot}}
\end{equation}



In the case relevant to the main text---$\phi = 0$, $\kappa_E\ll k_g$, and $k_E \ll k_\sigma$---both $R$ and $\alpha_\t{NL}$ can be expressed in terms of the non-inverted $\omega_0(\phi = 0) = \sqrt{(k_\sigma+ k_g)/I}\equiv \omega_+$ and inverted $\omega_0(\phi = \pi) = \sqrt{(k_\sigma- k_g)/I}\equiv \omega_+$ pendulum frequency, 
\begin{equation}
 R \approx \frac{\omega_+^2-\omega_-^2}{4g\omega_+} = \frac{8\alpha_\t{NL}}{g},
\end{equation}
highlighting a fundamental sensitivity-nonlinearity tradeoff.

\section{Frequency stability}
The frequency stability of a linear oscillator is fundamentally limited by thermal noise, modeled by the Allan deviation
\begin{equation}	\label{eq:ADthermal}
    \sigma_\omega^\text{th}(\tau) \approx \sqrt{\frac{\gamma}{2 \epsilon\tau}} 
\end{equation}
where $\gamma$ is the mechanical damping rate and $\epsilon = (\theta_0^2/2)/ \langle \theta_\t{th}^2\rangle $ is the ratio of the coherent displacement power to the thermal displacement power \cite{wang2021fundamental, wang2020frequency, sadeghi2020frequency, demir2021understanding, gavartin2013stabilization}.\footnote{Eq. \ref{eq:ADthermal} is derived in \cite{sadeghi2020frequency, demir2021understanding, wang2021fundamental, wang2020frequency, gavartin2013stabilization} assuming $\epsilon\gg 1$ and negligible readout noise. For a free-running oscillator and negligible readout noise, \cite{wang2020frequency,wang2021fundamental} show theoretically and experimentally that $\epsilon = 1$, while \cite{gavartin2013stabilization} show that $\epsilon = 1/2$.} Evidently, frequency noise can be reduced by coherently driving the oscillator. However, for a nonlinear oscillator, this advantage is countered by increased sensitivity to amplitude fluctuations (Eq. \ref{eq:freqShift}). Specifically, for small amplitude fluctuations $\sigma_{\theta_0}$ about a coherent drive amplitude $\theta_0$, one might anticipate 
\begin{equation}
    \sigma^\t{NL}_\omega  \approx \left|\frac{\partial \omega}{\partial \theta_0}\right|\sigma_{\theta_0} = 2 \omega_0 \alpha_\t{NL}\theta_0 \sigma_{\theta_0},
\end{equation}
implying the existence of a critical amplitude at which the advantage of increasing $\theta_0$ is saturated (Eq. 4 in the main text).

\vspace{-2mm}
\subsection{Frequency drift in free decay}
\vspace{-2mm}
Figure 3e shows the frequency drift of a micropendulum during an amplitude ringdown $\theta_0(t) = \theta_1 e^{-\gamma t/2}$, exhibiting an exponential decay due to amplitude-frequency coupling:
\begin{equation} \label{eq:freqShift}
        \omega(t)\approx \omega_0\left(1+\alpha_\text{NL} \theta_1^2 e^{-\gamma t}\right) \\
\end{equation}

Allan deviations for this type of drift, shown in Fig. 3f, exhibit a characteristic transition between $\sigma_\omega\propto \sqrt{1/\tau}$ (white noise) at short times and $\sigma_\omega\propto \tau$ at long times, with a minimum that depends on $\alpha_\t{NL}$, $\theta_1$ and the total measurement time $\tau_\t{m}$.  Here we derive an analytical expression for these curves.

Suppose an oscillator's frequency is measured for duration ${\tau_M}$ at sampling rate of ${{\Delta t}}^{-1}$. Each fractional frequency measurement $y_i=\omega_i/\omega_0$ has corresponding time $t_i = (i-1){\Delta t}$, where $i$ runs from $1$ to $M$ and $M= {\tau_M}/{\Delta t}$ is the number of measurements. The Allan variance as a function of averaging time $\tau$ can be estimated by averaging $n=\tau/{\Delta t}$ adjacent frequency measurements. The Allan variance (estimator) is 
\begin{equation} \label{eq:AVdef2}
	\hat{\sigma_y}^2 (n{\Delta t}) = \frac{1}{2n(M-2n+1)}\sum_{j=1}^{\frac{M-2n+1}{n}} \left(\sum_{i = nj}^{nj+n-1} y_{i+n} - y_i\right)^2. 
\end{equation}
During an amplitude ringdown $\theta_0(t)=\theta_1 e^{-\gamma t/2}$, assuming $\gamma \ll {{\Delta t}}^{-1}$ (high $Q$), ${\theta}_i\approx{\theta}(t_i)$ and Eq. \ref{eq:freqShift} can be written
\begin{equation} \label{eq:frequencyRingdown}
		y_i = 1 + \frac{\Delta \omega_1}{\omega_0} e^{-\gamma t_i},
\end{equation}
where ${\Delta \omega_i} \equiv \omega_i - \omega_0 = \omega_0 \alpha_\text{NL}{{\theta}_i}^2$ is the initial frequency shift. Plugging Eq. \ref{eq:frequencyRingdown} into the Eq. \ref{eq:AVdef2} yields
\begin{equation} \label{eq:ringdownAV}
    \begin{aligned}
        \hat{\sigma_y}^2 (\tau) = \left(\frac{{\Delta \omega_1}}{\omega_0}\right)^2 &\frac{{{\Delta t}}^2 e^{2\gamma {\Delta t} }}{2\tau ({\tau_M}-2 \tau + {\Delta t})} \times ... \\
        \times&\frac{\left(1 - e^{-\gamma \tau}\right)^4}{\left(e^{-\gamma {\Delta t}} -1\right)^2}  \frac{\left(1 - e^{-2 \gamma ({\tau_M}-2\tau +{\Delta t})}\right)}{e^{2\gamma \tau}-1} \\
    \end{aligned}
\end{equation}
On short intervals, the ringdown resembles a linear frequency drift $\Delta\omega_i\approx \Delta\omega_1(1-\gamma t_i)$ corresponds to an Allan deviation
\begin{equation}\label{eq:driftLinear} 
 \sigma_y^\t{NL}(\tau)\approx \frac{|\Delta\omega_1|}{Q\sqrt{2}}\tau
\end{equation}
as can be seen from Eq. \ref{eq:ringdownAV} assuming $\left\{\Delta t, \tau, \tau_M\right\}\ll \gamma^{-1}$.

The stability of the oscillator in free decay is therefore limited by both thermal noise (Eq. \ref{eq:ADthermal}), $\sigma_y\sim \sqrt{1/\tau}$ and amplitude drift (\ref{eq:ringdownAV}) $\sigma_y\sim \sqrt{\tau}$.  \textcolor{black}{This gives rise to a minimum
\begin{equation}
    \sigma_y\left(\tau_\text{min}\right) = \sqrt{\frac{3}{2}}\left(\frac{{\Delta \omega_1}}{2\omega_0 \epsilon Q^2}\right)^{1/3} = \sqrt{\frac{3}{2}}\left(\frac{{\alpha_\text{NL} {\theta_\text{th}}^2}}{ Q^2}\right)^{1/3}
\end{equation}
at averaging time
\begin{equation}
	\tau_\text{min} = \left(\frac{Q}{2\epsilon\omega_0{\Delta\omega_1}^2}\right)^{1/3} = \left(\frac{{\theta_\text{th}}^2 Q}{{\alpha_\text{NL}}^2 {\omega_0}^3 {\theta_1}^6}\right)^{1/3}.
\end{equation}}

	\renewcommand{\arraystretch}{1.2}
 \begin{table*}[t!]\color{black}
		\centering
		\begin{tabular}{llllllllllll}
			\Xhline{2\arrayrulewidth}
			&&& \hspace{1mm} & & & & \hspace{-15mm}Model (measurement) &\\
			\cline{5-10}
			Quantity &Symbol\;\;&Formula& \hspace{-2mm}$k_\sigma \gg k_{g,E}$\hspace{2mm} & Device 1 \;\;& Device 2\;\;& Device 3\;\; & Device 4 & Device 5$^*$\; & Device 6$^*$\; & Units\\
			\hline
			Suspension width& $w$ & & & 25 & 50 & 100 & 108 & 25 & 25 &  $\upmu\t{m}$\\
			\hspace{9.5mm} thickness& $h$ & & & 75 & 75 & 75 & 100 & 75 & 75 &   $\t{nm}$\\
			\hspace{9.5mm} \;\;\;\;length& $L$ & & & 7 & 7 & 7 & 7 & 7 & 7 & $\upmu\t{m}$\\
			\hspace{9.5mm} \;\;\;\;\;stress& $\sigma$ & & & 0.8 & 0.8 & 0.8 & 1.2 & 0.8 & 0.8 & $\t{GPa}$\\
			Paddle width& $w_\t{p}$ & & & 550 & 550 & 550 & 600 & 1100 & 550 &  $\upmu\t{m}$\\
			\hspace{9mm} thickness& $h_\t{p}$ & & & 95 & 95 & 95 & 400 & 95 & 190&  $\upmu\t{m}$\\
			Mass& $m$ & $\rho w_\t{p}^2 h_\t{p}$ & & 67 & 67 & 67 & 336& 268 & 134 & $\upmu\t{g}$\\
			Moment of inertia& $I$ & $\frac{m(w_p^2+4 h_p^2)}{12}$ & & 1.9 & 1.9 &  1.9 & 28 & 28 & 5.0 &  $\t{pg}\cdot \t{m}^2$ \\
			Stiffness - gravity& $k_g$ & $m g_0 h_\t{p}/2$ &  & 31 & 31 & 31 & 658 & 125 & 125 &  $\frac{\t{pN}\cdot\t{m}}{\t{rad}}$ \\
			\hspace{13mm}   tensile& $k_\sigma$ & $\sigma h w^3/3L$ & & 45 & 357 & 2900 & 6480 & 45 & 45 & $\frac{\t{pN}\cdot\t{m}}{\t{rad}}$  \\
			\hspace{13mm}   elastic& $k_E$ & $2 E h^3 w/3L$ & & 0.25 & 0.50 & 1.0 & 1.87 & 0.25 & 0.25 & $\frac{\t{pN}\cdot\t{m}}{\t{rad}}$\\
			\hspace{6mm}   nonlinearity\;\; & $k''_E$ & $3 E h w^5/5L^3$ & & 0.32 & 10 & 328 & 578 & 0.32 & 0.32 &  $\frac{\t{pN}\cdot\t{m}}{\t{rad}^3}$\\
			Frequency  & $\omega_0$ & $\sqrt{\frac{k_\sigma+k_g+k_E}{I}}$ & $\sqrt{\frac{k_\sigma+k_g}{I}}$ & 32 (32) & 72 (55) & 197 (157) & 80 (98) & 12 & 29 &  $2\pi\cdot\t{Hz}$\\
			\hspace{6.5mm} nonlinearity & $\alpha_\t{NL}$ & $\frac{1}{16}\frac{k''_E - k_g}{k_\sigma+k_g+k_E}$\hspace{2mm} & $\frac{1}{16}\frac{k''_E - k_g}{k_\sigma+k_g}$\hspace{2mm} &  -25 (20) & -3.3 (3) & 6.5 (6) & -0.7 (0.6) & -46 & -46 & $10^{-9}\frac{1}{\t{mrad}^{2}}$ \\
			Quality factor & $Q$ & $Q_0 \frac{k_\sigma+k_g+k_E}{k_E}$ & $Q_0 \frac{2\sigma h^2}{E w^2}$ & 1.7 (2) & 4.0 (3) & 15 (10) & 24 (5) & 3.5 & 3.5 &  $10^6$ \\
			Gravity sensitivity & $R$ & $\frac{\omega_0}{2g} \frac{k_g}{k_\sigma+k_g+k_E}$& $\frac{\omega_0}{2g} \frac{k_g}{k_\sigma}$  & 6.3 (5.0) & 2.9 (2.9) & 1.1 (1.) & 3.7 (3.5) & 4.6 & 10.8 & $2\pi\cdot\frac{\t{Hz}}{g_0}$  \\
			\hspace{9.5mm} resolution& $\Delta g$ & $\frac{2g_0}{Q_0}\frac{k_E}{k_g}$& $\frac{4g_0}{Q_0}\frac{k_E}{m h_p}$ & 3.1 (3.2) & (6.5) & 12 (14) & 1.0 (5.6) & 0.77 & 0.77 & $10^{-6}g_0$  \\
			Thermal torque & $S_\tau^\t{th}(\omega_0)$ & $4 k_B T I\omega_0/Q$& $\frac{4 k_B T k_E }{Q_0\sqrt{k_\sigma I}}$ & 6.3 & 5.9 & 5.0 & 9.9 & 10 & 6.6 & $10^{-20}\frac{\t{N}\cdot{\t{m}}}{\sqrt{\t{Hz}}}$ \\
			\hspace{7mm} acceleration\;\; & $S_a^\t{th}(\omega_0)$ &  $4S_\tau^\t{th}/(m h_\t{p})^2$ & & 2.0 (2) & 1.9 & 1.6 & 0.15 & 0.81 & 0.53 & $10^{-9}\frac{g_0}{\sqrt{\t{Hz}}}$\\
			\Xhline{2\arrayrulewidth}
		\end{tabular}
		\caption{Formulas used to model torsion pendula with diamond-shaped Si paddles as in Fig. 1(a), assuming Si density $\rho = 2330\;\t{kg}/\t{m}^3$, Si$_3$N$_4$ elastic modulus $E = 250\;\t{GPa}$, and intrinsic $Q_0(h) = 7000\cdot (h/\t{nm})$~\cite{villanueva2014evidence}.  Device 1 is the device in Figs. 1-3 and green in Fig. 4.  Device 2 (3) is red (blue) in Fig. 4.  Device 4 (using a different fabrication method on a higher stress wafer) is magenta in Fig. 4.  Devices 5 and 6 are variations on Device 1 with larger paddle dimensions.  $^*$Note that these devices are unstable in the inverted configuration, since $k_\sigma < k_g$.}
		\vspace{-2mm}
	\end{table*}

\color{black}
\vspace{-2mm}
\section{Linear dynamics and thermal noise} 
\vspace{-2mm}

In the limit of small displacement, dynamics of a driven, damped torsion pendulum can be modeled as \cite{saulson1990thermal}
\begin{equation}\label{eq:linearEOM}
	I\ddot{\theta}+ k_\t{tot}(1+i\phi (\Omega)) \theta = \tau
\end{equation}
where $\tau$ is the driving torque and 
\begin{equation}
\phi(\Omega) = \frac{k_E}{k_\t{tot}}\phi_0(\Omega) = \frac{1}{Q}\frac{\phi_0(\Omega)}{\phi_0(\omega_0)} = \frac{\gamma}{\omega_0}\frac{\phi_0(\Omega)}{\phi_0(\omega_0)}
\end{equation}
is the effective loss tangent of the pendulum, related to the (in general freqency dependent) loss tangent of the suspension material $\phi_0(\Omega)$ via dissipation dilution factor $k_\t{tot}/k_E$~\cite{pratt2023nanoscale}.  

Equivalently, one can write \cite{saulson1990thermal}
\begin{equation}\label{eq:linearEOM}
	I\ddot{\theta}+ I\gamma(\Omega)\dot{\theta}+I\omega_0^2 \theta = \tau
\end{equation}
where 
\begin{equation}
    \gamma(\Omega) = \gamma(\omega_0)\frac{\omega_0}{\Omega}\frac{\phi_0(\Omega)}{\phi_0(\omega_0)} 
\end{equation}
is a frequency-dependent damping rate.

Setting $\gamma(\omega_0) = \gamma$ for simplicity, the single-sided thermal displacement power spectral density can be expressed as
\begin{subequations}\begin{align}
S_\theta^\t{th}(\Omega) & = |\chi(\Omega)|^2 S_\tau^\t{th}(\Omega)\\
& \approx \frac{4 k_B T \gamma/I}{(\Omega^2 -\omega_0^2)^2 - \gamma^2 \omega_0^2}\frac{\omega_0}{\Omega}\frac{\phi_0(\Omega)}{\phi_0(\omega_0)} 
\end{align}\end{subequations}
where 
\begin{equation}
\chi(\Omega)\approx I^{-1}\left((\omega_0^2-\Omega^2)+i\gamma\omega_0\right)^{-1}
\end{equation}
is the mechanical susceptibility in the low $\phi$ limit and 
\begin{equation}
S_\tau^\t{th}(\Omega)\approx 4 k_B T I \gamma (\omega)
\end{equation}
is the thermal Langevin torque.

In Fig. 3(c), we model thermal noise spectra using the structural damping model $\phi(\Omega) = \phi(\omega_0)$

\subsection{Thermal acceleration noise}

Subjecting the torsion pendulum chip to a lateral acceleration $a$ produces an inertial torque
\begin{equation}
    \tau = m a r_\t{CM}
\end{equation}
where $m$ is the mass of the torsion paddle and $r_\t{CM}$ is the distance from the center-of-mass of the paddle to the torsion axis (denoted $r$, in Sec. 1).  It follows that displacement produced by acceleration noise is given by 
\begin{equation}\label{eq:displacementproducedbyacceleration}
    S_\theta(\Omega) = (m r_\t{CM})^2 |\chi(\Omega)|^2 S_a(\Omega)
\end{equation}
and that the acceleration equivalent thermal noise (thermal acceleration sensitivity) is given by 
\begin{equation}\label{eq:thermalacceleration}
S_a^\t{th}(\Omega) =  S_\tau^\t{th}(\Omega)/(m r_\t{CM})^{2}
\end{equation}

Equations \ref{eq:displacementproducedbyacceleration}- \ref{eq:thermalacceleration} are used to analyze data in Fig. 2(c,d).

\section{Geometry dependent device properties}

An extended version of Table 1 is provided in Table S1. In addition to devices 1-4 presented in the main text, we highlight two unstable devices with wider and thicker paddles.


 \color{black}

\color{black}
\vspace{-2mm}
\section{Experiment}
\vspace{-2mm}

In this section we provide various details on the experiment.
\color{black}

\vspace{-2mm}
\subsection{Device images}
\vspace{-2mm}

\begin{figure}[h!]
    \includegraphics[width=0.9\columnwidth]{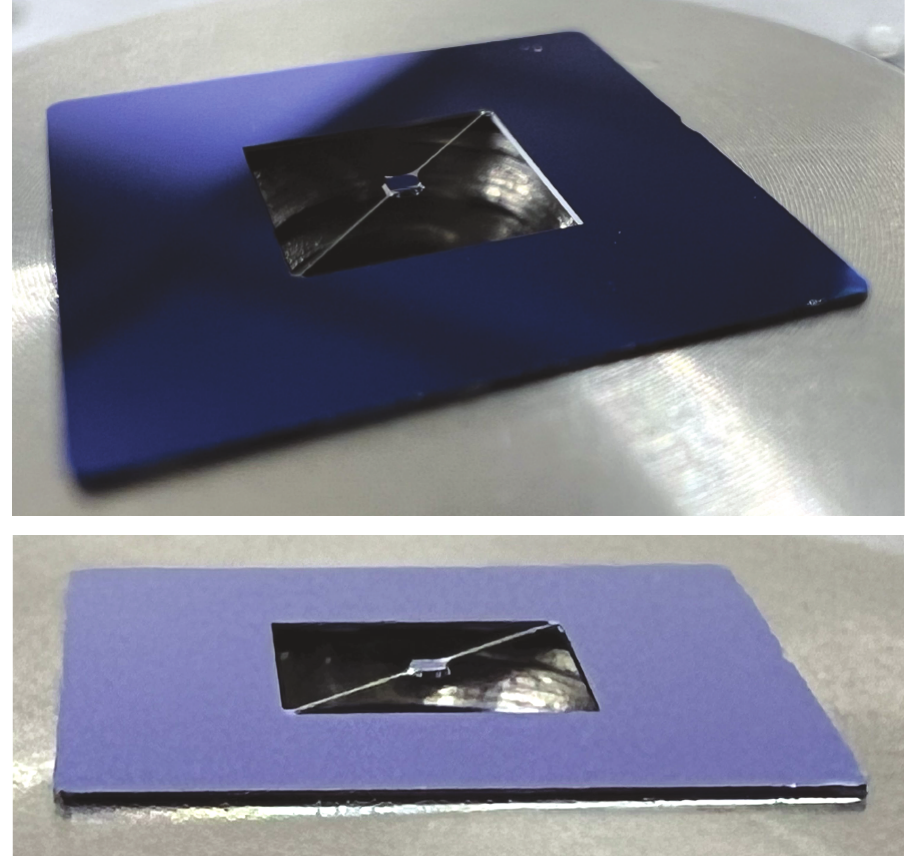}
			\caption{Additional images of the device shown in Fig. 1b.}
			\label{fig:S1}
\end{figure}

Additional images of the torsion micropendulum device shown in Fig. 1b are shown in Fig. \ref{fig:S1}.  The central torsion pad is suspended by 25-$\upmu\t{m}$-wide, 75-nm-thick Si$_3$N$_4$ ribbon suspensions spanning a $5\,\t{mm} \times 5\,\t{mm}$ window inside of a $12\,\t{mm} \times 12\,\t{mm}$ wide, 200-$\upmu$m-thick Si chip. Visible defects in the pad geometry are due to the non-uniformity of the KOH etch used to release the suspension. (Details of the fabrication procedure are described in the appendix of \cite{pratt2023nanoscale}.)  We have devised a new fabrication method employing deep reactive ion etching to more precisely define the pad dimensions (important for tailoring $\alpha_\t{NL}$); however, we have found that the KOH-etched devices reported here and in \cite{pratt2023nanoscale} have similar $Q$.

\vspace{-2mm}
\subsection{Frequency Sensitivity to Temperature}
\vspace{-2mm}

\begin{figure}[h!]
    \includegraphics[width=0.85\columnwidth]{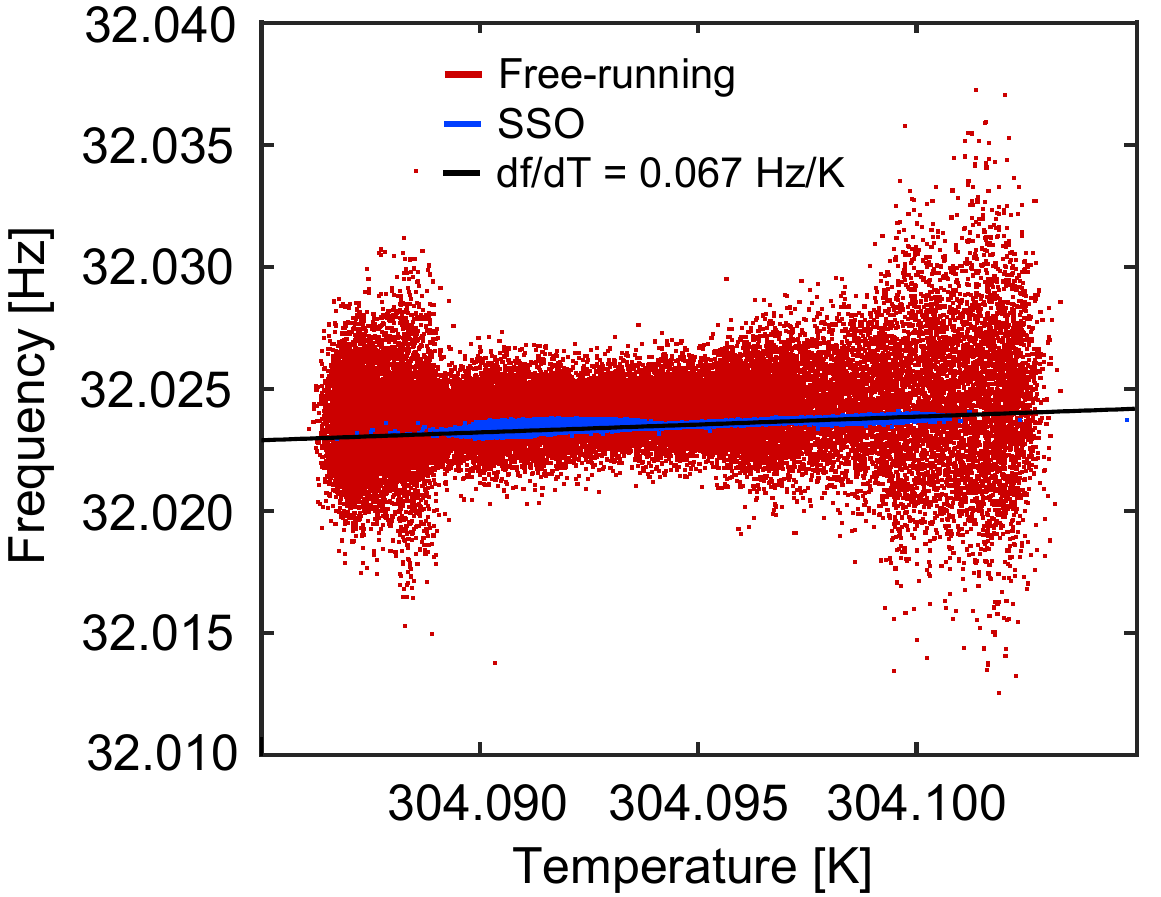}
			\caption{Parametric plot of frequency $f$ versus temperature $T$ for free-running and self-sustained oscillator (SSO) measurements in Fig. 3. Linear fit to the SSO and free-runnings measurements yield $df/dT = 0.067$ Hz/K and 0.070 Hz/K, respectively.} 
   \label{fig:S2}
    \vspace{0mm}
\end{figure}

To determine the sensitivity of the micropendulum's frequency to temperature $d f/d T$, frequency and temperature time series for both free-running (Fig. 3a) and self-sustained (Fig. 3g) oscillators were plotted against one another and fit to a line, as shown in Fig. \ref{fig:S2}. The fitted slopes yield $d f/d T = 0.070$ Hz/K and $0.067$ Hz/K, respectively.  

\begin{figure}[b!]
    \includegraphics[width=0.9\columnwidth]{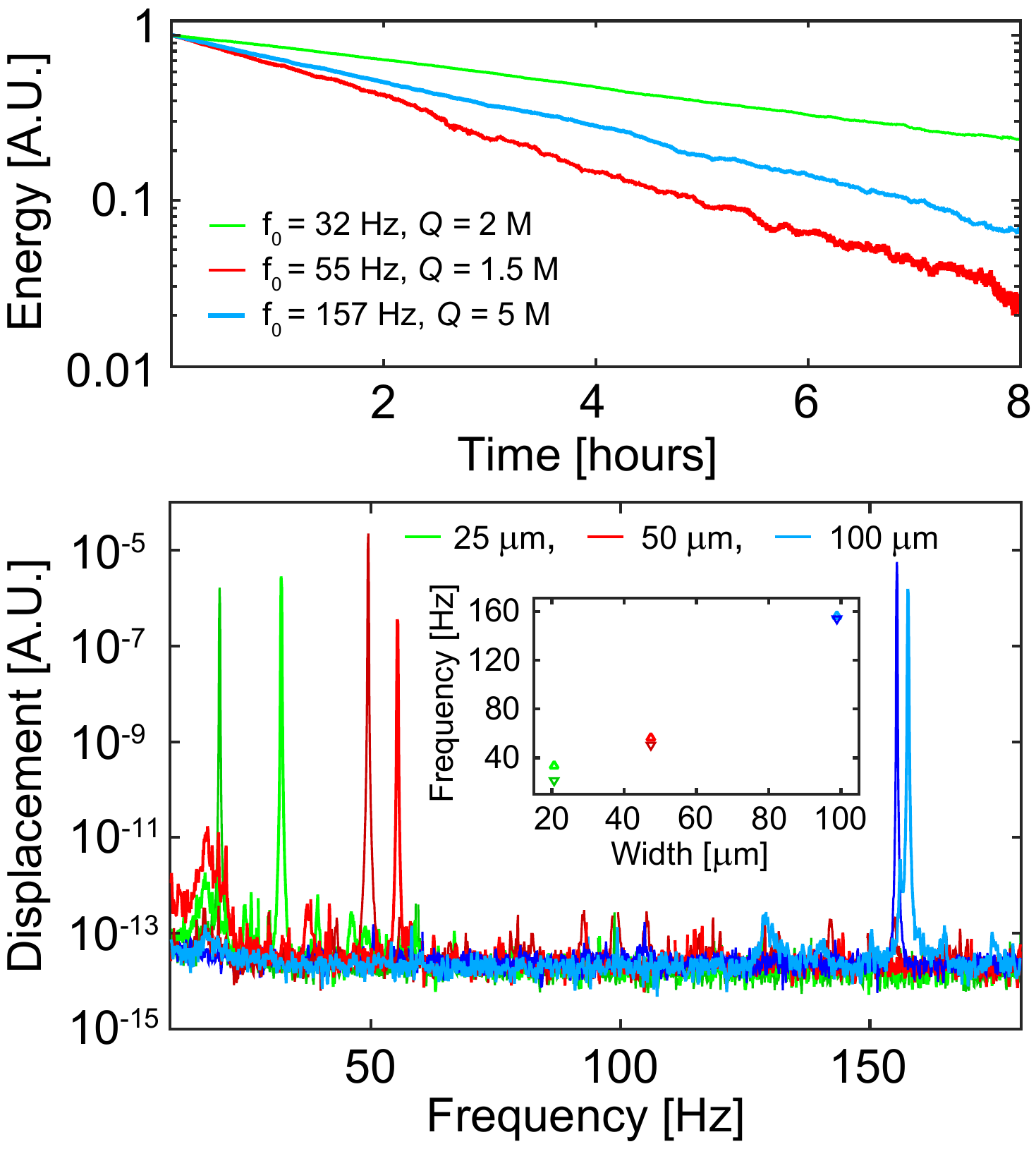}
			\caption{(top) Energy ringdowns of micropendula with 25, 50, and 100 $\mu$m wide suspensions shown in Fig. 4 in the main text. Quality factors here reflect current measurements following prolonged exposure to air and contaminants in storage and transport. (bottom) Displacement spectra of each in their normal (light colors) and inverted (dark colors) configuration. Inset: Frequency of each pendulum in both configurations, as a function of ribbon suspension width.}
			\label{fig:S3}
\end{figure}

\vspace{0mm}
\subsection{Nonlinear stiffness versus suspension width}
\vspace{-2mm}

Figure 4 in the main text shows frequency versus amplitude measurements for devices with three different ribbon suspension widths, $w$. The $w=50\,\upmu \t{m}$ and $w=100\,\upmu \t{m}$ devices are the same as studied in ~\cite{pratt2023nanoscale}. 
The $w=25\,\upmu\t{m}$ device is a new sample fabricated for this study.  To obtain the sensitivity estimates in Fig. 4b, we record the $Q$ and the inverted frequency of each device via ringdown and thermal noise spectra, respectively, as shown in Fig. \ref{fig:S3}.  The $Q$ factors shown in Fig. \ref{fig:S3}a have degraded relative to their original values \cite{pratt2023nanoscale}, likely due to contamination while in storage or transit between vacuum chambers.  We therefore use the original values in Fig. 4b.

As discussed in the main text, we have devised a strategy to balance the pendulum and suspension nonlinearity while preserving gravitational sensitivity, by carefully controlling the dimensions of the torsion paddle and ribbon suspensions.  A new fabrication process employing deep reactive ion etching (see below) has allowed us to realize a $\omega_0\approx 2\pi\cdot 100$ Hz micropendulum  with $\alpha_{\t{NL}}\approx 10^{-3}$ and $R\approx 5 \t{Hz}/g_0$ (see Table I).  The limits of this process are currently being explored.




\begin{figure}[t!]
    \includegraphics[width=0.5\columnwidth]{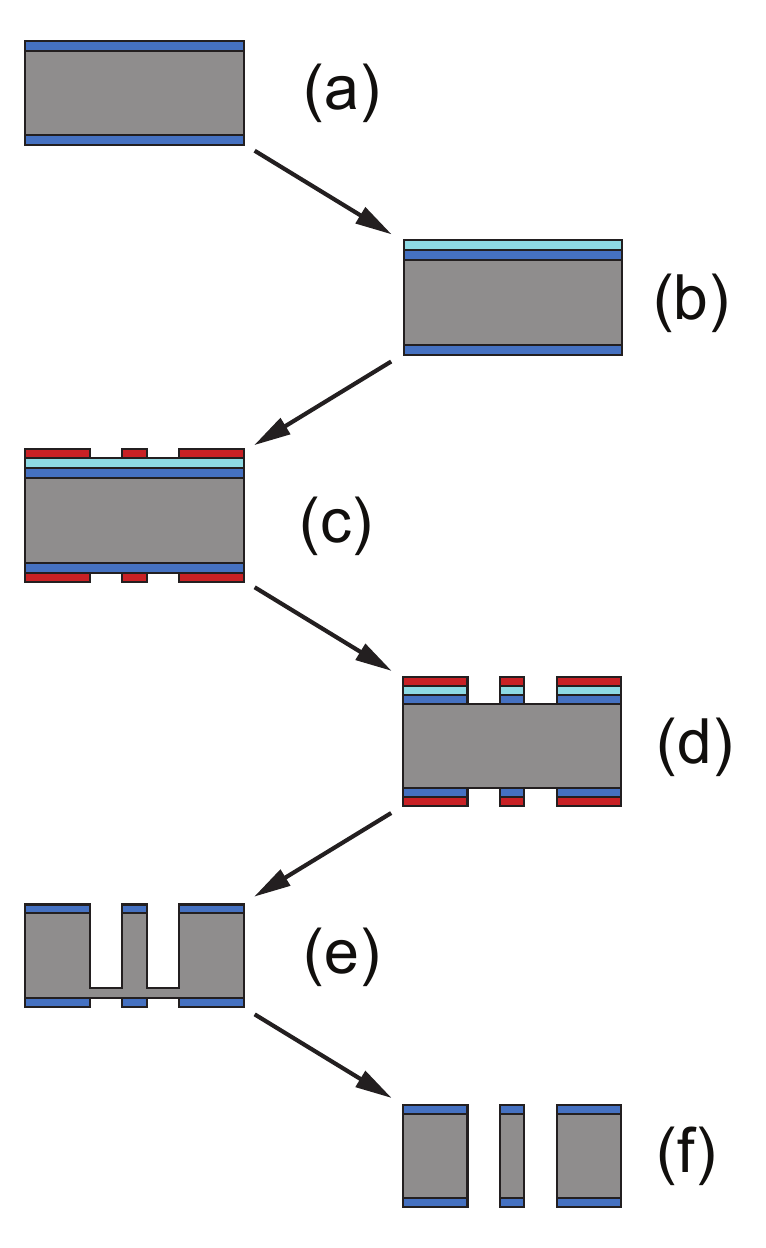}
			\caption{\color{black} Fabrication method for isochronous micropendulum. (a) A 100-nm-thick LPCVD Si$_3$N$_4$ thin film is deposited on each side of a silicon wafer. (b) A 1 $\upmu$m layer of PECVD SiO$_2$ is deposited on one side of the wafer to serve as a hard mask for DRIE. (c) A layer of photoresist is used on each side for photolithography patterning. (d) The device pattern is transferred to the wafer by etching away the oxide and nitride layers. (e) A deep silicon etch is used to remove all but $10\, \upmu$m of the bulk silicon. (f) KOH is used to etch the remaining Si, releasing the micropendulum from its frame. } 
   \label{fig:S5}
    \vspace{0mm}
\end{figure}

\color{black}
\vspace{-2mm}
\subsection{Fabrication of isochronous micropendulum}
\vspace{-2mm}

The new fabrication method for our isochronous device is outlined in Fig. \ref{fig:S5}.  It uses deep reactive-ion etching (DRIE) to preserve the rectangular geometry of the Si paddle and avoid irregularities associated with the selectivity of KOH etching against different crystal planes in Si. This allows us to fabricate paddles whose mass and moment of inertia better match our model. In turn, this enables us to predict the width of the ribbon suspensions necessary to balance their nonlinear stiffness against the pendulum nonlinearity, minimizing $\alpha_\t{NL}$. 
\color{black}

\newpage

\bibliography{ref}